\documentclass[12pt]{article}
\usepackage{amssymb,amsmath,epsfig}
\allowdisplaybreaks

\begin{document}

\title{\bf Warm Modified Chaplygin Gas Shaft Inflation}
\author{Abdul Jawad \thanks{jawadab181@yahoo.com;~~abduljawad@ciitlahore.edu.pk}, Amara
Ilyas\thanks{amara{\_}Ilyas14@yahoo.com}~
and Shamaila Rani \thanks {shamailatoor.math@yahoo.com, drshamailarani@ciitlahore.edu.pk}\\
Department of Mathematics, COMSATS Institute of\\ Information
Technology, Lahore-54000, Pakistan.}

\date{}
\maketitle
\begin{abstract}
In this paper, we examine the possible realization of a new family
of inflation called ``shaft inflation" by assuming the modified
Chaplygin gas model and tachyon scalar field. We also consider the
special form of dissipative coefficient as
$\Gamma={a_0}\frac{T^{3}}{\phi^{2 }}$ and calculate the various
inflationary parameters in the scenario of strong and weak
dissipative regimes. In order to examine the behavior of
inflationary parameters, the planes of $n_s - \phi,~n_s - r$ and
$n_s - \alpha_s$ (where $n_s,~\alpha_s,~r$ and $\phi$ represent
spectral index, its running, tensor to scalar ratio and scalar field
respectively) are being developed which lead to the constraints: $r<
0.11$, $n_s=0.96\pm0.025$ and $\alpha_s =-0.019\pm0.025$. It is
quite interesting that these results of inflationary parameters are
compatible with BICEP$2$, WMAP $(7+9)$ and recent Planck data.
\end{abstract}

\section{Introduction}

Inflation is the most acceptable paradigm that describes the physics
of the very early universe. Besides of solving most of the
shortcomings of the hot big-bang scenario, like the horizon, the
flatness and the monopole problems \cite{R1}-\cite{Linde:1983gd},
inflation also generates a mechanism to explain the large-scale
structure (LSS) of the universe \cite{R2}-\cite{R205} and the origin
of the anisotropies observed in the cosmic microwave background
(CMB) radiation \cite{astro}-\cite{Ade:2015lrj}. The primordial
density perturbations may be soured from quantum fluctuations of the
inflaton scalar field during the inflationary expansion. The
standard cold inflation scenario is divided into two regimes: the
slow-roll and reheating phases. In the slow-roll period, the
universe undergoes an accelerated expansion and all interactions
between the inflaton scalar field and other field's degrees of
freedom are typically neglected. Subsequently, a reheating period
\cite{Kofman:1994rk, Kofman:1997yn} is invoked to end the brief
acceleration. After reheating, the universe is filled with
relativistic particles and thus the universe enters in the radiation
big-bang epoch. For a modern review of reheating, see
\cite{Amin:2014eta}.

On the other hand, warm inflation is an alternative mechanism for
having successful inflation. The warm inflation scenario, as opposed
to standard cold inflation, has the essential feature that a
reheating phase is avoided at the end of the accelerated expansion
due to the decay of the inflaton into radiation and particles during
the slow-roll phase \cite{warm1,warm2}. During warm inflation, the
temperature of the universe did not drop dramatically and the
universe can smoothly entered into the decelerated,
radiation-dominated period, which is essential for a successful
big-bang nucleosynthesis. In the warm inflation scenario,
dissipative effects are important during the accelerated expansion,
so that radiation production occurs concurrently with the
accelerated expansion. The dissipative effect arises from a friction
term $\Gamma$ which describes the processes of the scalar field
dissipating into a thermal bath via its interaction with other
field's degrees of freedom.

The effectiveness of warm inflation may be parameterized by the
ratio $R\equiv \Gamma/3H$. The weak dissipative regime for warm
inflation is for $R\ll 1$, while for $R\gg1$, it is the strong
dissipative regime. Following
Refs.\cite{Zhang:2009ge,BasteroGil:2012cm}, a general
parametrization of the dissipative coefficient depending on both the
temperature of the thermal bath $T$ and the inflaton scalar field
$\phi$ can be written as
\begin{equation}
\Gamma(T,\phi)=C_{\phi}\,\frac{T^{m}}{\phi^{m-1}}, \label{G}%
\end{equation}
where  the parameter $C_{\phi}$ is related with the dissipative
microscopic dynamics, the exponent $m$ is an integer whose value is
depends on the specifics of the model construction for warm
inflation and on the temperature regime of the thermal bath.
Typically, it is found that $m=3$ (low temperature), $m=1$ (high
temperature) or $m=0$ (constant dissipation).

Later on, Linde \cite{a13} introduced the chaotic inflation by
realizing that the initial conditions for scaler field driving
inflation which may help in solving the persisting inflationary
problems. A plethora of works in the subject of warm inflation along
with chaotic potential have been done. For instance, Herrera
\cite{a2a} investigated the warm inflation in the presence of
chaotic potential in loop quantum cosmology and found consistencies
of results with observational data. Del campo and Herrera
\cite{9sad} discussed the warm inflationary model in the presence of
standard scalar field, dissipation coefficient of the form
$\Gamma\propto \phi^{n}$ and generalized Chaplygin gas (GCG) and
extract various inflationary parameters. Setare and Kamali
investigated warm tachyon inflation by assuming intermediate
\cite{2tsad} and logamediate scenarios \cite{1tsad}. Bastero-Gill et
al. \cite{6sad} obtained the expressions for the dissipation
coefficient in supersymmetric (SUSY) models and their result
provides possibilities for realization of warm inflation in SUSY
field theories. Bastero-Gill et al. \cite{5sad} have also explored
inflation by assuming the quartic potential. Herrera et al.
\cite{9tsad} studied intermediate inflation in the context of GCG
using standard and tachyon scalar field.

Panotopoulos and Vidaela \cite{i17} discussed the warm inflation by
assuming quartic potential and decay rate proportional to
temperature and found that the results of inflationary parameters
are compatible with the latest Planck data. Moreover, many authors
have investigated the warm inflation in various alternative as well
as modified theories of gravity \cite{aj,bam1}.
Recently, a new family of
inflation models is being developed named as shaft inflation
\cite{shaft}. The idea of this inflation was that the inflationary
flatness is effected by shaft i.e; when the scalar field found
itself nearest to one of them, it slow-rolls inside the shaft, until
inflation ends and gives way to the hot big bang cosmology. The
generalized form of shaft potential is
\begin{equation}\label{s}
V(\phi)=\frac{M_p^4 \phi ^{2n-2}}{\left(\phi
^n+m^n\right)^{2-\frac{2}{n}}}\,,
\end{equation}
where $M_p, m, n$ are massless constants.

In the present, we discuss the warm inflation by assuming shaft
potential, modified Chaplygin gas model and tachyon scalar field. We
will extract the inflationary parameters. The formate of the paper
is as follows: In the next section, we will discuss the detailed
inflationary scenario with tachyon field and generalized dissipative
coefficient. Section \textbf{3} and \textbf{4} contains the
information about disordered parameters for shaft potential in
strong and weak dissipative regimes, respectively. In section
\textbf{5}, the results are given in summarized form.

\section{Tachyon Scalar Field Inflationary Scenario}

The universe undergoes an accelerated expansion of the universe The
responsible for this acceleration of the late expansion is an exotic
component having a negative pressure, usually known as dark energy
(DE). Several models have been already proposed to be DE candidates,
such as cosmological constant \cite{de1}, quintessence
\cite{de2}-\cite{de4}, k-essence \cite{de5}-\cite{de7}, tachyon
\cite{de8}-\cite{de10}, phantom \cite{de11}-\cite{de13}, Chaplygin
gas \cite{de14}, holographic dark energy \cite{Li:2004rb}, among
others in order to modify the matter sector of the gravitational
action. Despite the plenty of models, the nature of the dark sector
of the universe, i.e. DE and dark matter, is still unknown. There
exists another way of understanding the observed universe in which
dark matter and DE are described by a single unified component.
Particularly, the Chaplygin gas \cite{de14} achieves the unification
of DE and dark matter. In this sense, the Chaplygin gas behaves as a
pressureless matter at the early times and like a cosmological
constant at late times. The original Chaplygin gas is characterized
by an exotic equation of state with negative pressure
\begin{equation}
p=-\frac{\beta}{\rho},\label{ocg}
\end{equation}
whit $\beta$ being a constant parameter. The original Chaplygin gas
has been extended to the so-called generalized Chaplygin gas (GCG)
with the following equation of state \cite{gcg}
\begin{equation}
p_{gcg}=-\frac{\beta}{\rho^{\sigma}},\label{ocg}
\end{equation}
with $0\leq \sigma \leq 1$. For the particular case $\lambda=1$, the
original Chaplygin gas is recovered. The main motivation for
studying this kind of model comes from string theory. The Chaplygin
gas emerges as an effective fluid associated with D-branes which may
be obtained from the Born-Infeld action \cite{bi}. At background
level, the GCG is able to describe the cosmological dynamics
\cite{Makler:2002jv}, however the model presents serious issues at
perturbative level \cite{Amendola:2003bz}. Thus, a modification to
the GCG, results in the modified Chaplygin gas (MCG) with an
equation of state given by \cite{Benaoum:2002zs}
\begin{equation}
p=\omega\rho-\frac{\beta}{\rho^{\sigma}},\label{mcg}
\end{equation}
where $\omega$ is a constant parameters with $0\leq \sigma\leq1$, is
suitable to describe the evolution of the universe
\cite{Lu:2008zzb,Debnath:2004cd} which is also consistent with
perturbative study \cite{SilvaeCosta:2007xy}.

The energy conservation equation for MCG
model turns out to be
\begin{equation}\label{c1}
\rho_{mcg}=\bigg(\frac{\beta}{1+\omega}+\frac{\upsilon}{a^{3(1+\sigma)(1+\omega)}}\bigg)^{\frac{1}{1+\sigma}}\,,
\end{equation}
where $\upsilon$ is constant of integration. In spatially flat FRW
model, Friedmann equation described as
\begin{eqnarray}\label{jj}
H^2&=&\frac{1}{3 M_p^2}(\rho_{m} + \rho_{\gamma})\,,
\end{eqnarray}
where $\rho_{m}$ is the energy density of matter field and
$\rho_{\gamma}$ is the energy density of radiation field. The warm
MCG model modifies first Friedmann equation which has been used in
Eq.(\ref{jj}) reduces to
\begin{eqnarray}\label{jf}
H^2&=&\frac{1}{3M_p^2}\bigg[\bigg(\frac{\beta}{1+\omega}+\upsilon
\rho_{\phi}^{(1+\sigma)(1+\omega)}\bigg)^{\frac{1}{1+\sigma}}+\rho_{\gamma}\bigg]\,,
\end{eqnarray}
which is named as Chaplygin gas inspired inflation. The energy
density and pressure of tachyon scalar field are defined as follows
\cite{i16},
\begin{equation}\label{b}
\rho_{\phi}=\frac{V(\phi)}{\sqrt{1-{\dot{\phi}}^2}}\,,\quad
p_{\phi}=-V(\phi)\sqrt{1-{\dot{\phi}}^2}\,.
\end{equation}
The inflaton and imperfect fluid energy densities according to the
Eq. (\ref{b}) are conserved as
\begin{eqnarray}\label{o}
&&\dot{\rho_\phi}+3H(\rho_\phi+p_\phi)=-\Gamma\dot{\phi}^2\,,
\\\label{o2}
&&\dot{\rho_{\gamma}}+4H\rho_\gamma=\Gamma\dot{\phi}^{2}\,,
\end{eqnarray}
where $\Gamma$ is the dissipation factor that evaluates the rate of
decay of $\rho_\phi$ into $\rho_\gamma$. It is also important to
note that this decay rate can be used as a function of the
temperature of the thermal bath and the scalar field, i.e.,
$\Gamma(T, \phi)$ or a function of only temperature of thermal bath
$\Gamma(T)$, or a function of scalar field only $\Gamma(\phi)$, or
simply a constant.

The second law of thermodynamics indicates that $\Gamma$ must be
positive, so the inflaton energy density decompose into radiation
density. The second conservation equation is given by
\begin{eqnarray}\nonumber
\frac{\ddot{\phi}}{1-\dot{\phi}^2}+3H\dot{\phi}+\frac{V'(\phi)}{V(\phi)}&=&
-\frac{\Gamma\dot{\phi}}{V}\sqrt{1-{\dot{\phi}}^2}\,,
\\\nonumber\Rightarrow~~~\ddot{\phi}+\bigg(3H+\frac{\Gamma}{V}\bigg)\dot{\phi}
&=&-\frac{V'(\phi)}{V(\phi)}\,,~~~~~~~~~~~~~~~~~~~~~\dot{\phi}\ll{1}
\\\label{g}\Rightarrow~~~ 3H(1+R)\dot{\phi}&=&-\frac{V'(\phi)}{V(\phi)}\,,
~~~\text{where}~~~\ddot{\phi}\ll(3H+\frac{\Gamma}{V})\dot{\phi}\,,
\end{eqnarray}
where $R=\frac{\Gamma}{3 H V}$. In weak dissipative epoch, $R\ll1$
runs to $\Gamma\ll3H$ while $R\gg1$ indicates the high dissipative
regime. Here, we assume some constraints which leads to static
epoch, i.e., $\rho_\phi\approx V(\phi)$, slow-roll limit,
$V(\phi)\gg \dot{\phi}^{2}$, $(3H+\Gamma)\dot{\phi}\gg\ddot{\phi}$,
quasi-stable decay of $\rho_\phi$ into $\rho_\gamma$,
$4H\rho_\gamma\gg\dot{\rho_\gamma}$ and $\Gamma \dot{\phi}^2\gg
\dot{\rho_\gamma}$. As we know that the energy density of scalar
field is much greater than the energy density of radiations but also
at the same time, the energy can be larger than the expansion rate
with $\rho_{\gamma}^\frac{1}{4}> H$. This is approximately equal to
$T > H$  by considering thermalization, which is true condition to
take place in warm inflation. With the help of all these limits Eqs.
(\ref{jj}), (\ref{o2}) and (\ref{g}) become
\begin{eqnarray}\label{y}
H^2&=&\frac{1}{3M_p^2}\bigg(\frac{\beta}{1+\omega}+\upsilon
\rho_{\phi}^{(1+\sigma)(1+\omega)}\bigg)^{\frac{1}{1+\sigma}}\,,\\\label{yy}4H\rho_\gamma&=&\Gamma
\dot{\phi}^2\,,\\\label{yyy}3H(1+R)\dot{\phi}&=&-\frac{V'(\phi)}{V(\phi)}\,,
\end{eqnarray}
where prime represents the derivative with respect to $\phi$.

The energy density of radiation can be used as $C_\gamma T^4$ when
we have taken the thermalization. Here $C_\gamma=\pi^2 g_*/30$,
where $g_*$ shows the degree of freedom. This expression gives the
value as $C_\gamma \simeq 70$ with $g_* = 228.75$. The temperature
of thermal bath can be obtained by merging the Eqs. (\ref{yy}) and
(\ref{yyy})
\begin{equation}\label{i}
T=\left(\frac{\Gamma {V'}^2}{36 C_\gamma H^3 V^2
(1+R)^2}\right)^\frac{1}{4}\,,
\end{equation}
where $\Gamma={a_0}\frac{T^{q}}{\phi^{{q}-1}}$, which is the general
form of dissipative coefficient, while $a_0$ and $q$ are constant
parameters associated with dissipative microscopic dynamics. The
consequences of radiation are studied during inflation through this
kind of dynamic which is suggested first time in warm inflation with
theoretical basis of supersymmetry (SUSY) \cite{i24,i25}.

A set of dimensionless slow-roll parameters must be satisfied for
the occurrence of warm inflation which are defined in the form of
Hubble parameter as \cite{i26}
\begin{equation*}
\epsilon = -\frac{\dot{H}}{H^2}, \quad \eta =
-\frac{\ddot{H}}{H\dot{H}}\,.
\end{equation*}
The slow-roll parameters can also be deduced in the form of scalar
field and thermalization according to the tachyon field along with
modified Chaplygin gas, which are defined as
\begin{eqnarray}\nonumber
\epsilon &=&
\frac{\upsilon  (1+\omega )M_p^2 V^{(1+\sigma ) (1+\omega )-2}V'^2}{2(1+R)
\left(\frac{\beta }{1+\omega }+\upsilon  V^{(1+\sigma ) (1+\omega )}\right)^{1+\frac{1}{1+\sigma }}}\,,
\\\nonumber \eta &=& \frac{M_p^2}{(1+R)\left(\frac{\beta }{1+\omega }+\upsilon  V^{(1+\sigma )
(1+\omega )}\right)^{\frac{1}{1+\sigma }}}\left[\frac{((1+\sigma )
(1+\omega
)-2)V'^2}{V^2}+\frac{2V''}{V}\right]\\\nonumber&-&2(1+\sigma
)\epsilon \,.
\end{eqnarray}
We can describe number of e-folds in terms of Hubble parameter as
well as inflaton
\begin{equation}\label{hhh}
N(\phi)= \int^{t_{f}}_{t_{i}} H dt = \int^{\phi_{f}}_{\phi_{i}}
\frac{H}{\dot{\phi}} d\phi\,,
\end{equation}
where $\phi_{i}$ and $\phi_{f}$ can be calculated with the help of
first and second slow-roll parametric conditions, i.e.,
$\epsilon=1+R$ and $|\eta|=1+R$.

Next, we will calculate some inflationary parameters such as tensor
and scalar power spectra $(P_R,P_g)$, tensor and scalar spectral
indices $(n_R, n_s)$. The form of scalar power spectrum can be
estimated as $P_R(k_0)\equiv\frac{25}{4}\delta_H^2(k_0)$, where
density disorders $\delta_H^2(k_0)\equiv\frac{k_F (T_R)}{2\pi^2}$
and $k_F =\sqrt{\Gamma H}$. However, the amplitude of the tensor and
scalar power spectrum of the curvature perturbation are given by
\begin{eqnarray}\label{p}
P_R&\simeq&
\left(\frac{H}{2\pi}\right)^2\left(\frac{3H^2}{V'}\right)^2
\left(\frac{T}{H}\right)\left(1+R\right)^\frac{5}{2}, \quad P_g
\simeq24 \kappa \left(\frac{ H}{2 \pi}\right)^2\,.
\end{eqnarray}
The tensor-to-scalar ratio can be computed by using the relation $r=
\frac{P_R}{P_g}$. However, the spectral index and running of
spectral index are defined as \cite{i28}
\begin{eqnarray}\label{hh}
n_s &=&1 +\frac{d\ln{P_R}}{d\ln{k}},\quad \alpha_s=\frac{d
{n_s}}{d\ln{k}}\,.
\end{eqnarray}
Here, the interval in wave number $k$ is referred to the number of
e-folds $N$, through the expression as
\begin{equation}\label{hn}
d\ln{k}= - d{N}\,.
\end{equation}
In the following, we will evaluate the inflationary parameters for
weak and strong dissipative regimes.

\section{Inflationary Parameters in Strong Epoch with Shaft Potential}

The special case of shaft potential where $n=2$ is considered for
which Eq.(\ref{s}) takes the form as $V(\phi)=\frac{M_p^4
\phi^{2}}{\left(\phi^2+m^2\right)}$. The temperature of the
radiation for present model with the help of shaft potential, Eq.
(\ref{i}) takes the following form
\begin{eqnarray}\label{j}
T &=&\left(\frac{m^6 M_p^5 \sqrt{3\left(\frac{\beta }{1+\omega
}+\upsilon \left(\frac{M_p^4 \phi ^2}{m^2+\phi
^2}\right)^{(1+\omega) (1+\sigma )}\right)^{-\frac{1}{1+\sigma
}}}}{a_0 \left(m^2+\phi ^2\right)^3 \left(M_p^4-\frac{M_p^4 \phi
^2}{m^2+\phi ^2}\right) C_{\gamma }}\right)^{\frac{1}{7}}\,.
\end{eqnarray}
The number of e-folds can be calculated by Eq. (\ref{hhh}) with
$\dot{\phi}=\frac{-V'}{3 H V R}$ for strong regime as
\begin{eqnarray}\nonumber
N&=&\frac{a_0^{\frac{4}{7}}}{2 3^{2/7} m^{\frac{2}{7}}
M_p^{\frac{4}{7}}} \int _{\phi _i}^{\phi
_f}\left(\sqrt{\left(\frac{\beta }{1+\omega }+\upsilon
\left(\frac{M_p^4 \phi ^2}{m^2+\phi ^2}\right)^{(1+\omega) (1+\sigma
)}\right)^{\frac{1}{1+\sigma
}}}\right)^{\frac{10}{7}}\\\label{ne}&\times& \bigg(\frac{
\left(m^2+\phi ^2\right)^{\frac{1}{7}}}{C_{\gamma }^{3/7}\phi
}\bigg)d\phi\, .
\end{eqnarray}
For the strong epoch, $\phi_i$ and $\phi_f$ can be described by
considering $\epsilon=R$ and $|\eta|=R$ respectively. The power
spectrum attains the value from Eq.(\ref{p}) as follows
\begin{eqnarray}\nonumber
P_R&=&\frac{a_0^{10/7} \left(\frac{m^2 M_p^4}{m^2+\phi
^2}\right)^{10/7} }{48 3^{3/14}\pi ^2 C_{\gamma
}^{15/14}}\left(\frac{\left(\frac{\beta }{1+\omega }+\upsilon
\left(\frac{M_p^4 \phi ^2}{g^2+\phi ^2}\right)^{(1+\omega )
(1+\sigma )}\right)^{\frac{1}{2 (1+\sigma
)}}}{M_p}\right)^{10/7}\\\nonumber&\times&\left(\frac{\left(\frac{m^2
M_p^4 \phi }{\left(m^2+\phi ^2\right)^2}\right)^{1/7} }{
m^{20/7}\left(\frac{M_p^4 \phi ^2}{m^2+\phi
^2}\right)^{25/7}}\right) \,.
\end{eqnarray}
The scalar power spectrum is given by
\begin{equation}\label{hg}
P_g=\frac{2}{9\pi  M_p^4}\left(\frac{\beta }{1+\omega }+\upsilon
V^{(1+\sigma ) (1+\omega )}\right)^{\frac{1}{1+\sigma }}\,.
\end{equation}
The tensor-to-scalar ratio can be found by using expression
(\ref{hg}) which yields
\begin{eqnarray}\nonumber
r&=& \frac{32m^2 M_p^5 \upsilon  \phi  \left(\frac{m^2 M_p^4 \phi
}{\left(m^2+\phi ^2\right)^2}\right){}^{15/7} \left(\frac{M_p^4 \phi
^2}{m^2+\phi ^2}\right)^{\frac{3}{7}+\omega +\sigma +\omega \sigma
}3^{2/7} (1+\omega ) a_0 C_{\gamma }^{3/7}}{\left(\frac{a_0 m^4
M_p^4}{m^2+\phi ^2}\right)^{11/7}\left(\upsilon
\left(\frac{M_p^4\phi ^2}{m^2+\phi ^2}\right)^{(1+\sigma ) (1+\omega
)}+\frac{\beta }{1+\omega }\right)^{1+\frac{1}{2 (1+\sigma )}}
}\\\nonumber&\times&\left(\frac{\left(\upsilon  \left(\frac{M_p^4
\phi ^2}{m^2+\phi ^2}\right)^{(1+\sigma ) (1+\omega )}+\frac{\beta
}{1+\omega }\right)^{\frac{1}{2+2 \sigma }}}{M_p}\right)^{3/7} \,.
\end{eqnarray}
Figure \textbf{1} shows the plot of tensor-to-scalar ratio versus
spectral index within strong regime. This ratio is being plotted for
three different values of $m$ with the condition $m<\phi$. Red line
has been plotted for $m=0.2$, green dashed line for $m=0.5$ and blue
dotted line for $m=0.9$. According to the plot, ratio is not
satisfied with spectral index when $m=0.2$ while tensor-to-scalar
ratio is compatible with the spectral index for other two values.
\begin{figure}
\centering {\includegraphics [bb=10 12 270 200, clip=true,
width=7cm, height=7cm, keepaspectratio]{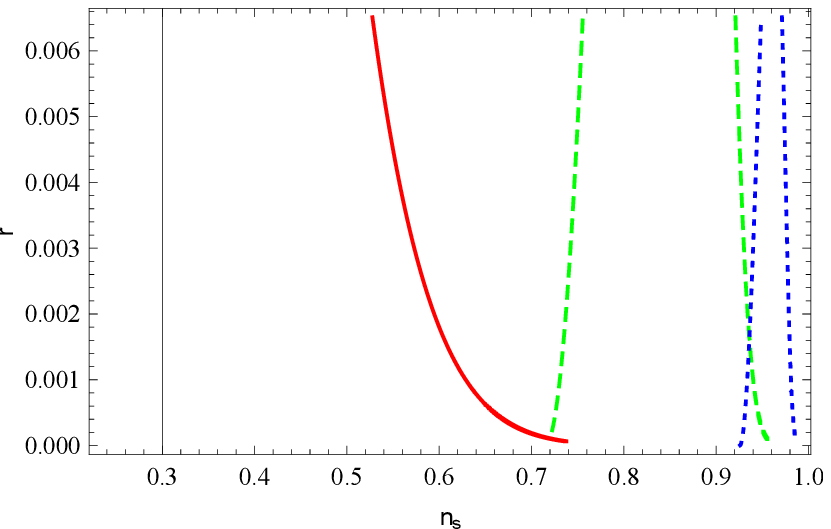}} \caption{Plot of
tensor-to-scalar ratio verses spectral index in strong epoch with
$a_0 = 2\times10^6$.}
\end{figure}

However, the spectral index and it's running attained the values by
using Eqs. (\ref{hh}) and (\ref{hn}) as
\begin{eqnarray}\nonumber
n_s&=&1-\left(\frac{2 3^{2/7} \phi ^3 C_{\gamma }^{3/7} M_p^8
\left(\frac{m^4 a_0 M_p^4}{m^2+\phi ^2}\right){}^{3/7} }{7
\left(m^2+\phi ^2\right)^5 a_0 \left(\frac{m^2 \phi
M_p^4}{\left(m^2+\phi ^2\right)^2}\right)^{13/7} \left(\frac{\beta
}{1+\omega }+\upsilon  \left(\frac{\phi ^2 M_p^4}{m^2+\phi
^2}\right)^{(1+\sigma ) (1+\omega )}\right)
}\right)\\\nonumber&\times&\frac{\left(\frac{\beta  \left(49 m^2+23
\phi ^2\right)}{1+\omega }-\upsilon  \left(-23 \phi ^2+m^2 (-39+10
\omega )\right) \left(\frac{\phi ^2 M_p^4}{m^2+\phi
^2}\right){}^{(1+\sigma ) (1+\omega )}\right)}{\left(\frac{\phi ^2
M_p^3 \left(\frac{\beta }{1+\omega }+\upsilon  \left(\frac{\phi ^2
M_p^4}{m^2+\phi ^2}\right){}^{(1+\sigma ) (1+\omega
)}\right)^{\frac{1}{2+2 \sigma }}}{m^2+\phi ^2}\right)^{4/7}}\,.
\end{eqnarray}
We plot spectral index $n_s$ versus  scalar field $\phi$ in Figure
\textbf{2} and notice that red line which represents the behavior of
spectral index with respect to $\phi$ for $m=0.2$ requires a very
large value of $\phi$ to reach in the range of spectral index. The
other two different values i.e., $m=0.5$ and $m=0.9$ with green and
blue lines respectively, satisfies the range of spectral index for
$\phi\in[1, 50]$. It can be observed that the tensor-to-scalar ratio
($r$) remains less than $0.11$ for the range of spectral index $0.96
< n_s < 0.97$ in the strong dissipative epoch.
\begin{figure}
\centering {\includegraphics [bb=10 12 370 200, clip=true,
width=7cm, height=7cm, keepaspectratio]{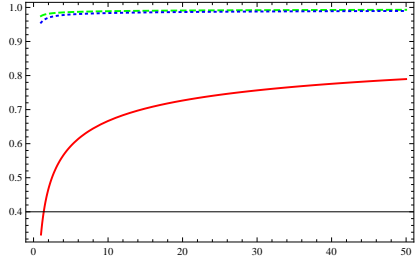}} \caption{Plot of
spectral index number w.r.t inflaton in strong epoch with $a_0 =
2\times10^6$.}
\end{figure}

The running of spectral index becomes
\begin{eqnarray}\nonumber
\alpha_s&=&-\bigg[8 3^{4/7} \phi  C_{\gamma }^{6/7} \left(\frac{m^2
\phi M_p^4}{\left(m^2+\phi ^2\right)^2}\right)^{9/7} \left(\frac{m^4
a_0 M_p^4}{m^2+\phi ^2}\right)^{6/7} \bigg(\left(\frac{\beta
}{1+\omega }\right)^2 \big(231 m^2 \phi ^2\\\nonumber&+&23 \phi
^4\big)+2 \frac{\upsilon \beta }{1+\omega }\big(23 \phi ^4+m^2 \phi
^2 (214-17 \omega )+7 m^4 (1+\omega ) (12+5 \omega +5
\sigma\\\nonumber&\times& (1+\omega ))\big) \left(\frac{\phi ^2
M_p^4}{m^2+\phi ^2}\right)^{(1+\sigma ) (1+\omega )}+\upsilon ^2
\big(23 \phi ^4+m^2 \phi ^2 (197-34 \omega )-2
m^4\\\nonumber&\times& (1+\omega ) (-39+10 \omega )\big)
\left(\frac{\phi ^2 M_p^4}{m^2+\phi ^2}\right)^{2 (1+\sigma )
(1+\omega )}\bigg)\bigg]\left(49 m^{10} a_0^2 M_p^4\right)^{-1}
\bigg(\frac{\beta }{1+\omega }\\\nonumber&+&\upsilon
\left(\frac{\phi ^2 M_p^4}{m^2+\phi ^2}\right)^{(1+\sigma )
(1+\omega )}\bigg)^{-2} \left(\frac{\phi ^2 M_p^3 \left(\frac{\beta
}{1+\omega }+\upsilon \left(\frac{\phi ^2 M_p^4}{m^2+\phi
^2}\right)^{(1+\sigma ) (1+\omega )}\right)^{\frac{1}{2+2 \sigma
}}}{m^2+\phi ^2}\right)^{-8/7}\,.
\end{eqnarray}
The plot of running of spectral index  with respect to scalar field
is shown in Figure \textbf{3}. The suggested values for running of
spectral index by WMAP$7$ \cite{i29, i30} and WMAP$9$ \cite{i31} are
approximately equal to $-0.992 \pm 0.019$ and $-0.019 \pm 0.025$,
respectively. It can be observed that this parameter is compatible
with observational data for $m=0.5$ and $m=0.9$. However for
$m=0.2$, the plot of running of spectral index is not compatible
with the required range of spectral index.
\begin{figure}
\centering {\includegraphics [bb=10 12 270 200, clip=true,
width=7cm, height=7cm, keepaspectratio]{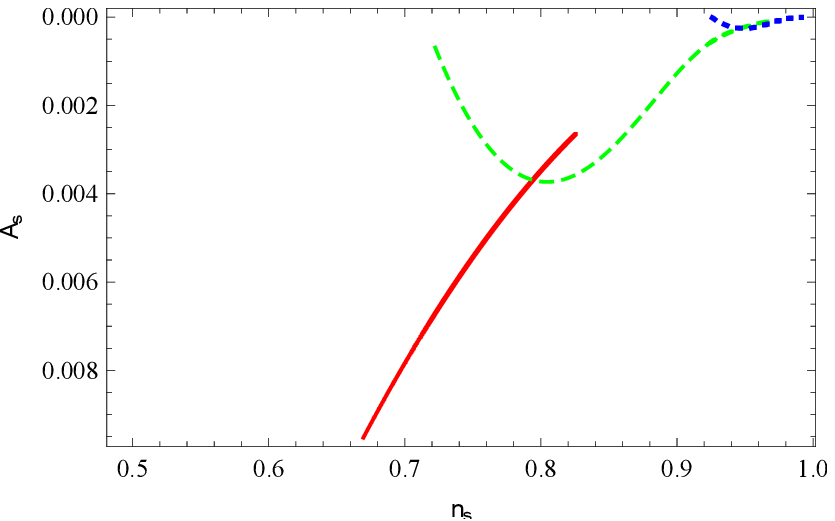}} \caption{Plot for
running of spectral index versus spectral index in strong epoch with
$a_0 = 2\times10^6$.}
\end{figure}

\section{Inflationary Parameters in Weak Epoch with Shaft Potential}

Here we study the tachyon model in weak epoch $(R\ll1)$, the
temperature of the radiation for present model with the help of
shaft potential, Eq. (\ref{j}) takes the form as
\begin{eqnarray}\label{uj}
T &=&\frac{a_0 m^4 M_p^3 \left(\frac{\beta }{1+\omega }+\upsilon
\left(\frac{\phi ^2 M_p^4}{m^2+\phi ^2}\right)^{(1+\sigma )
(1+\omega )}\right)^{-\frac{3}{2 (1+\sigma )}}}{\sqrt{3} \phi ^4
\left(m^2+\phi ^2\right)^2 C_{\gamma }}\,.
\end{eqnarray}
The number of e-folds can be calculated by Eq. (\ref{hhh}) with
$\dot{\phi}=\frac{-V'}{3 H V}$ as
\begin{eqnarray}\nonumber
N&=&\frac{1}{2 m^2 M_p^2}\int _{\phi _i}^{\phi _f}\left(\frac{\beta }
{1+\omega }+\upsilon  \left(\frac{\phi ^2 M_p^4}{m^2+\phi ^2}
\right)^{(1+\sigma ) (1+\omega )}\right)^{\frac{1}{1+\sigma }}
\left(m^2+\phi ^2\right) \phi d\phi  \, ,
\end{eqnarray}
where $\phi_{i}$ and $\phi_{f}$ can be found by taking $\epsilon =1$
and $|\eta|=1$ respectively.

The power spectrum attains the value from Eq.(\ref{p}) as
\begin{eqnarray}\nonumber
P_r&=&\frac{\left(m^2+\phi ^2\right)^3 a_0 } {48 m^2 \pi ^2 \phi ^6
C_{\gamma } M_p^{14}}\left[M_p^4-\frac{\phi ^2 M_p^4}{m^2+\phi
^2}\right]\left[\frac{\beta }{1+\omega }+\upsilon \left(\frac{\phi
^2 M_p^4}{m^2+\phi ^2}\right)^{(1+\sigma ) (1+\omega
)}\right]^{\frac{1}{1+\sigma }}\,.
\end{eqnarray}
The scalar power spectrum remains same as for the strong regime. The
tensor-to-scalar ratio is obtained by using expressions of power
spectrum and scalar spectrum, which is given by
\begin{eqnarray}\nonumber
r&=& \frac{8 M_p^2 (1+\omega ) \upsilon  }{\left(\frac{M_p^4 \phi
^2}{m^2+\phi ^2}\right)^{2-(1+\omega ) (1+\sigma )}} \left(\frac{2
M_p^4\phi }{m^2+\phi ^2}-\frac{2 M_p^4 \phi ^3}{\left(m^2+\phi
^2\right)^2}\right)^2\\\nonumber &\times&\left(\frac{\beta
}{1+\omega }+\upsilon  \left(\frac{M_p^4 \phi ^2}{m^2+\phi
^2}\right)^{(1+\omega) (1+\sigma )}\right)^{-1-\frac{1}{1+\sigma
}}\,.
\end{eqnarray}
Figure \textbf{4} shows the plot of tensor-to-scalar ratio versus
spectral index within weak regime. Tensor-to-scalar ratio is being
plotted for three different values of $m$ with the condition
$m<\phi$. Red line has been plotted for $m=0.2$, green dashed line
for $m=0.5$ and blue dotted line for $m=0.9$. According to the plot,
there is no change in the behavior of tensor-to-scalar ratio for
spectral index while tensor -to-scalar ratio is compatible with the
spectral index for all values of $m$.
\begin{figure}
\centering {\includegraphics [bb=10 12 270 200, clip=true,
width=7cm, height=7cm, keepaspectratio]{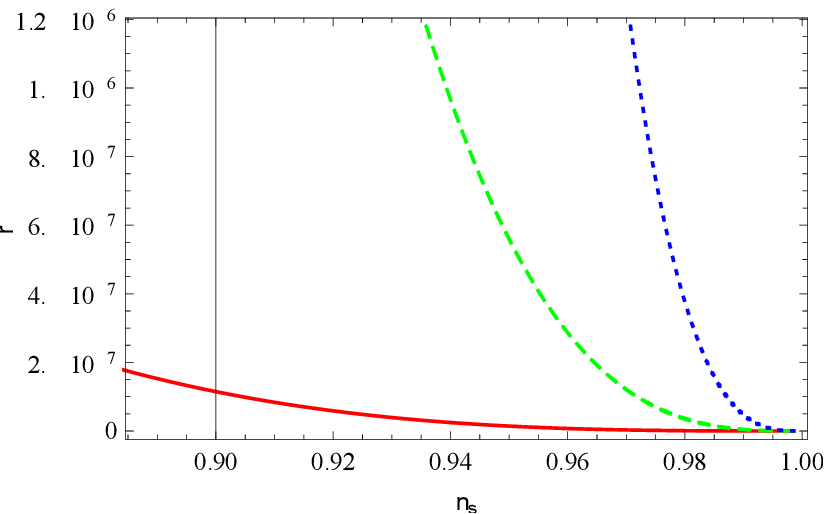}} \caption{Plot of
tensor-to-scalar ratio verses spectral index in weak epoch with $a_0
= 2\times10^6$.}
\end{figure}

The value of spectral index is found with the help of above motioned
power spectrum along with first part of Eq.(\ref{hh}) and
Eq.(\ref{hn}). It is given as follows
\begin{eqnarray}\nonumber
n_s &=&1+\left(\frac{\beta }{1+\omega }+\upsilon \left(\frac{M_p^4
\phi ^2}{m^2+\phi ^2}\right)^{(1+\omega ) (1+\sigma
)}\right)^{\frac{-1}{1+\sigma }}\bigg[\frac{8 m^6 M_p^{10} (1+\omega
) \upsilon  \phi }{\left(m^2+\phi
^2\right)^5}\\\nonumber&\times&\left(\frac{M_p^4 \phi ^2}{m^2+\phi
^2}\right)^{(1+\omega ) (1+\sigma )}\left(\frac{\beta }{1+\omega
}+\upsilon \left(\frac{M_p^4 \phi ^2}{m^2+\phi ^2}\right)^{(1+\omega
) (1+\sigma )}\right)^{-1}\\\label{ns}&-& \frac{4
M_p^2}{\left(m^2+\phi ^2\right)^3}\left(m^4 \left(1+2 M_p^4\right)+2
m^2 \phi ^2+\phi ^4\right)\bigg]\,.
\end{eqnarray}
Figure \textbf{5} represents the spectral index versus scalar field
for $m=0.2,~m=0.5$ and $m=0.9$. According to WMAP$7$ \cite{i29,
i30}, WMAP$9$ \cite{i31} and Planck $2015$ \cite{i32}, the value of
spectral index lies in the ranges $0.967 \pm 0.014,~0.972 \pm 0.013$
and $ 0.968 \pm 0.006$.
\begin{figure}
\centering {\includegraphics [bb=10 12 370 200, clip=true,
width=7cm, height=7cm, keepaspectratio]{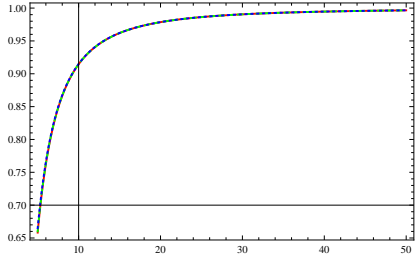}} \caption{Plot of
spectral index number w.r.t inflaton in weak epoch with $a_0 =
2\times10^6$.}
\end{figure}

Using Eqs.(\ref{ns}) and (\ref{hn}), the running of spectral index
is calculated as
\begin{eqnarray}\nonumber
\alpha_s&=&\frac{32 m^4 M_p^8}{\phi  \left(m^2+\phi ^2\right)^{11}}
\left(\frac{\beta }{1+\omega }+\upsilon \left(\frac{M_p^4 \phi
^2}{m^2+\phi ^2}\right)^{(1+\sigma ) (1+\omega )}\right)^{\frac{-2
(2+\sigma )}{1+\sigma }} \bigg[\left(\frac{\beta }{1+\omega
}\right)^2 \\\nonumber&\times&\phi ^2 \left(m^2+\phi ^2\right)^4
\left(m^4 \left(1+6 M_p^4\right)+2 m^2 \phi ^2+\phi
^4\right)+\left(\frac{M_p^4\phi ^2}{m^2+\phi ^2}\right)^{-1+\sigma
+\omega +\sigma  \omega } \\\nonumber&\times&\big(2 \phi
^{10}+m^{10} \left(1+2 M_p^4\right) (1+\omega )+m^2 \phi ^8
(9+\omega )+4 m^4 \phi ^6 \big(4+3 M_p^4\\\nonumber&+&\omega
\big)+m^8 \phi \big(6 \phi +M_p^8(1+\omega ) (3+2 \omega +2 \sigma
(1+\omega ))+4 \phi  \big(\omega +M_p^4 (4\\\nonumber&+&\omega
)\big)\big)+m^6\phi ^3 \left(-9 M_p^8(1+\omega )+2 \phi \left(7+3
\omega +M_p^4(13+\omega )\right)\right)\big)\frac{\upsilon \beta
M_p^8 \phi ^4}{1+\omega }\\\nonumber&+&M^8 \upsilon ^2 \phi ^4
\left(\frac{M_p^4\phi ^2}{m^2+\phi ^2}\right)^{2 (\sigma +\omega
+\sigma  \omega )} \big[\phi ^{10}+m^{10} \left(1+2 M_p^4\right)
(1+\omega )+m^2 \phi ^8\\\nonumber&\times& (5+\omega )+2 m^4 \phi ^6
\left(5+3 M_p^4+2 \omega \right)+m^6 \phi ^3 \big(-9 M_p^8 (1+\omega
)+2 \phi \big(5\\\nonumber&+&3 \omega +M_p^4(7+\omega
)\big)\big)+m^8 \phi \big[-M_p^8 (1+\omega ) (1+2 \omega )+\phi
\big(5+4 \omega +2 M_p^4\\\nonumber&\times&(5+2 \omega
)\big)\big]\big]\bigg]\,.
\end{eqnarray}

The plot of running of spectral index with respect to scalar field
is shown in Figure \textbf{6}. It can be observed that the running
of spectral index is compatible with observational data for
$m=0.2,~m=0.5$ and $m=0.9$.
\begin{figure}
\centering {\includegraphics [bb=10 12 270 200, clip=true,
width=7cm, height=7cm, keepaspectratio]{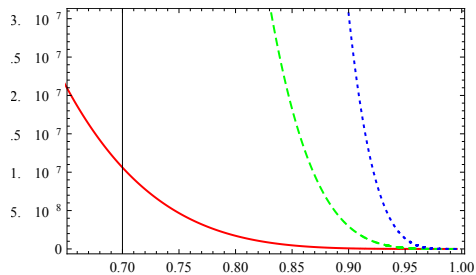}} \caption{Plot for
running of spectral index versus spectral index in weak epoch with
$a_0 = 2\times10^6$.}
\end{figure}

\section{Concluding Remarks}

The warm MCG inflationary scenario is being investigated with shaft
potential for tachyon scalar field. We have discussed this
inflationary scenario for both (weak and strong) dissipative regimes
in flat FRW universe. We have also examined the results for some of
necessary inflationary parameters such as the slow-roll parameters,
number of e-folds, scalar-tensor power spectra, spectral indices,
tensor-to-scalar ratio and running of scalar spectral index. We have
analyzed these parameters for strong epoch as well as weak regime by
using the special case of shaft potential. We have restricted
constant parameters of the models according to WMAP$7$ results for
examining the physical behavior of $n_s - \phi,~n_s - R$ and $n_s -
\alpha_s$ trajectories in both cases.

We have analyzed the behavior of inflationary parameters according
to two dimensionless parameters $(a_0,m)$ where the value of
$a_0=2\times10^6$ remains same for all necessary parameter. All the
trajectories are plotted for three different values i.e.,
$m=0.2,~m=0.5$ and $m=0.9$. The case for $m=0.2$ in strong epoch,
the plots showed the unsuitable behavior to satisfy the required
range of inflationary parameters. However, this value showed the
suitable behavior for weak regime. The standard values of parameters
are as: the tensor-to-scalar ratio $r<0.36,~ 0.38,~ 0.11$, the
spectral index $n_s=0.982\pm 0.020,~ 0.992\pm0.019,~
0.9655\pm0.0062$ according to WMAP$7$ \cite{i29, i30}, WMAP$9$
\cite{i31} and Planck 2015 \cite{i32} results respectively. In our
case, the tensor-to-scalar ratio versus spectral index is compatible
with this observational data (Figure \textbf{1} and \textbf{4}).
Also, Figure \textbf{3} and \textbf{6} clearly showed the
compatibility of spectral index for it's running with observational
data since observational values of running of spectral index are
$\alpha_{s}=-0.0084\pm0.0082,~-0.034\pm0.026,~ -0.019\pm0.025$
according to Planck $2015$ \cite{i32}, WMAP$7$ \cite{i29, i30} and
WMAP$9$ \cite{i31}, respectively.

\end{document}